# Bias-dependent Intrinsic RF Thermal Noise Modeling and Characterization of Single Layer Graphene FETs

Nikolaos Mavredakis, Anibal Pacheco-Sanchez, Paulius Sakalas, Wei Wei, Emiliano Pallecchi, Henri Happy, and David Jiménez

*Abstract*—In this article, the bias-dependence of intrinsic channel thermal noise of single-layer graphene field-effect transistors (GFETs) is thoroughly investigated by experimental observations and compact modeling. The findings indicate an increase of the specific noise as drain current increases whereas a saturation trend is observed at very high carrier density regime. Besides, short-channel effects like velocity saturation also result in an increment of noise at higher electric fields. The main goal of this work is to propose a physics-based compact model that accounts for and accurately predicts the above experimental observations in short-channel GFETs. In contrast to long-channel MOSFET-based models adopted previously to describe thermal noise in graphene devices without considering the degenerate nature of graphene, in this work a model for short-channel GFETs embracing the 2D material's underlying physics and including a bias dependency is presented. The implemented model is validated with de-embedded high frequency data from two short-channel devices at Quasi-Static region of operation. The model precisely describes the experimental data for a wide range of low to high drain current values without the need of any fitting parameter. Moreover, the consideration of the degenerate nature of graphene reveals a significant decrease of noise in comparison with the non-degenerate case and the model accurately captures this behavior. This work can also be of outmost significance from circuit designers' aspect, since noise excess factor, a very important figure of merit for RF circuits implementation, is defined and characterized for the first time in graphene transistors.

## I. Introduction

GRAPHENE field-effect transistors (GFETs) have been shown to exhibit significant extrinsic maximum oscillation ($f_{max}$) and unity-gain ($f_t$) frequencies with $f_t \sim 40$ GHz, $f_{max} \sim 46$ GHz for $SiO_2$ substrate devices [1] and $f_t \sim 70$ GHz, $f_{max} \sim 120$ GHz for devices with SiC substrate [2]. This promising performance despite the still early stage of the technology is mainly due to the extraordinary intrinsic characteristics of graphene, e.g., high carrier mobility and saturation velocity, leading circuit designers to consider these devices for analog RF applications whereas the lack of bandgap makes GEFTs unsuitable for digital circuitry [3]. Among such analog RF circuits [4] – [8], a Low Noise Amplifier (LNA) [7], [8] is a key circuit for receiver front-end systems. Hence, understanding of High Frequency Noise (HFN) in short-channel GFETs is of great importance and shall be modelled precisely.

In this study, we focus on intrinsic channel drain current noise, generated from the local random thermal fluctuations of the charge carriers, resulting to velocity fluctuations ($<v^2>$) and diffusion noise. For the two-port noise representation, the channel thermal noise fluctuations should be calculated as drain current noise spectral density ($S_{ID}$). Under Quasi-Static (QS) conditions quite below $f_t$ [9], channel thermal noise is independent of frequency. First works on a thermal noise analysis in FETs were reported several decades ago [10]-[12], whereas a long-channel compact model was first proposed by *Tsividis* [13, (8.5.21)]. Short-channel related effects were shown to increase $S_{ID}$ [14] and to account for this, physics-based compact models embracing Velocity Saturation (VS) effect were developed for CMOS devices [15]-[21].

A limited number of works dealing with the HFN characteristics of GFETs is available in the literature [22]-[27]. In order to improve the understanding of noise behavior, and enhance further the technology, a reliable description of $S_{ID}$ is required. Up to now, simple long-channel empirical models taken from MOSFETs, are used to describe $S_{ID}$ in fabricated GFETs [23], [24], [27] which neither consider the degenerate nature of graphene and its effect on noise [28]-[30], nor the behavior of noise at different operating conditions. Thus, the main objective of this study is the development of a physics-based compact model for $S_{ID}$ of single-layer (SL) GFETs which accounts both for the noise bias dependence including the VS effect and the degenerate nature of graphene. The approach presented here is based on an already established chemical-potential based model, describing the GFET IV, small-signal and 1/f noise characteristics [31]-[33]. To our knowledge, this

N. Mavredakis, A. Pacheco-Sanchez and D. Jiménez are with the Departament d'Enginyeria Electrònica, Escola d'Enginyeria, Universitat Autònoma de Barcelona, Bellaterra 08193, Spain. (e-mail: Nikolaos.mavredakis@uab.es).
P. Sakalas is with the MPI AST Division, 302 Dresden, Germany, also with the Semiconductor Physics Institute of Center for Physical Sciences and Technology, LT-10257, Vilnius, Lithuania, and also with the Baltic Institute for Advanced Technologies, LT-01403, Vilnius, Lithuania.
W. Wei, E. Pallecchi and H. Happy are with Univ. Lille, CNRS, UMR 8520 - IEMN, F-59000 Lille, France.

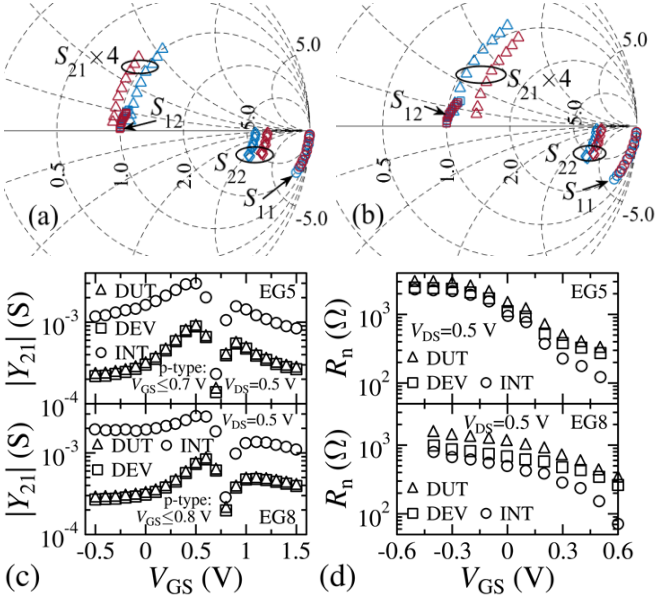

Fig. 1. Measured (DUT) S-parameters from 0.8 GHz to 8.4 GHz at $V_{GS}$=-0.5 V (blue) and 0.5 V (red) for GFETs with a) L=200 nm (EG5) and b) L=300 nm (EG8). DUT, de-embedded (DEV) and intrinsic (INT-after the removal of $R_G$, $R_c$ effect) magnitude of c) $Y_{21}$ parameter and d) noise resistance $R_n$ vs. $V_{GS}$ at 1 GHz for both EG5, EG8 GFETs. P-type operation region for $V_{GS} \leq 0.7$ V (EG5) and 0.8 V (EG8). All data in (a)-(d) reported at $V_{DS}$=0.5 V.

is the first time that such a complete $S_{ID}$ model is proposed and validated with experimental data of short-channel GFETs [34] without the need of any fitting parameter, after appropriate de-embedding procedures for both Y- parameters [35], [36] and noise data [37], [38]. In Section II, the devices under test (DUT) and the HFN measurement set-up are described in detail while in Section III, the derivation of the $S_{ID}$ model is presented thoroughly. Finally, in Section IV, the behavior of both the model and experiments vs. bias is presented where apart from the Power Spectral Density (PSD) of noise, the significant - from circuit designers' point of view-excess noise factor parameter γ [12], [39] and the intrinsic noise resistance $R_{nINT}$ [40] are also shown for the first time in GFETs.

## II. DUT AND MEASUREMENT SET-UP

Two SL short-channel aluminum back-gated CVD GFETs fabricated on a 300 nm thick $SiO_2$ followed by 40 nm Al deposition and lift-off process, were characterized in this study with a ~4 nm thick $Al_2O_3$ used as a dielectric layer between graphene and gate. The total width was W=12x2 μm=24 μm (where 2 is the number of gate fingers) and gate length L=200 nm (EG5) and L=300 nm (EG8), respectively. More details on GFET fabrication can be found elsewhere [34]. On-wafer DC and AC standard characteristics (S (Y)$_{DUT}$) have been measured with a PNA-X N5247A and a Keysight HP4142 Semiconductor Parameter analyzer. Noise parameters (HFN$_{DUT}$) in source-load matching conditions have been measured using the corrected Y-factor technique with a Maury Microwave automated tuner system ATS 5.21 and impedance tuner MT982. Gate voltage $V_{GS}$ was swept from strong p-type to strong n-type region whereas drain voltage was set to the maximum limit for the specific GFETs, $V_{DS}$=0.5 V. Operation frequency for noise measurements was set to 1 GHz as the main goal of this work is to study $S_{ID}$ at the QS regime ($f_T$ ~9 GHz and $f_T$ ~4 GHz for EG5 and EG8, respectively [34]). A complete de-embedding procedure was applied to both Y-parameters and HFN data [35]-[38] (see Supplementary Information SI, §A1). Pad parasitic network de-embedding from HFN$_{DUT}$ and Y$_{DUT}$ yields device data: HFN$_{DEV}$ and Y$_{DEV}$. Since the basic goal of this work is to express intrinsic channel thermal noise, the effect of contact and gate resistances $R_c$, $R_G$, respectively, should also be excluded from HFN$_{DEV}$ and Y$_{DEV}$ parameters since $R_c$ is not negligible in GFETs [41]. After this removal, intrinsic HFN$_{INT}$ and Y$_{INT}$ are obtained [36, (10)-(13)], (see SI, §A2). To extract intrinsic $S_{ID}$ from HFN parameters data, the following relation is used [16, (16)], [21, (5)]:

$$S_{ID} = 4 \cdot K_B T_0 |Y21_{INT}|^2 R_{nINT} \quad (1)$$

where $K_B$ is the Boltzmann constant and $T_0$ is the standard reference (290 K) [16] temperature. Notice that only $Y_{21INT}$ and $R_{nINT}$ are required to extract $S_{ID}$ from experimental data. Fig. 1a-1b present the measured $S_{DUT}$ parameters in a Smith chart at $V_{GS}$=-0.5, 0.5 V and $V_{DS}$=0.5 V for frequencies from 0.8 to 8.4 GHz for EG5 (a) and EG8 (b) GFETs whereas, Fig. 1c-1d depict |$Y_{21DUT, DEV, INT}$| and $R_{nDUT, DEV, INT}$, respectively vs. $V_{GS}$ for the same DUTs and $V_{DS}$ where the contribution of $R_c$, $R_G$ (EG5: $R_G$=18 Ω, EG8: $R_G$=12 Ω) to $R_{nINT}$ is significant.

To ensure that the IV model [31], [32] describes accurately the DC operating point at HF operation, the former is validated with $\Re[Y_{21DEV}]$, i.e., the transconductance $g_m$ of the device measured through the HF set-up. A model parameter embracing defects effects and the initial state of traps at different lateral fields [42], i.e. a trap-induced hysteresis, has been considered here. Trap-affected performance of the technology used here has been described elsewhere [42] using the same model. Fig. 2 presents the modeled and measured $g_m$ for both GFETs under test vs. $V_{GS}$ at $V_{DS}$=0.5 V. The model agreement with $\Re[Y_{21DEV}]$ is precise, especially in p-type region, cf. Fig. 2. In this study we focus on p-type region due to maximum $g_m$ recorded there since data asymmetries [33] are observed between p- and n-type regions cf. Fig. 1c, Fig. 2. P-type region is defined for $V_{GS} \leq 0.7$ and $\leq 0.8$ V for EG5, EG8, respectively as shown in Fig. 1c. $g_m$ data, obtained from the derivative of drain current $I_D$ w.r.t $V_{GS}$ matches $g_m$ (AC) as shown in Fig. 2. The extracted model parameters for both GFETs are listed in the Table I where $\mu$ is the carrier mobility, $C_{back}$ the back-gate capacitance, $V_{BSO}$ the flat-band voltage, $R_c$ the contact resistance, $\Delta$ - the inhomogeneity of the electrostatic potential, which is related to the residual charge density $\rho_0$, $\hbar\Omega$ is the phonon energy related to VS effect and $K_{tr}$ reveals the $V_{DS}$ dependence of the trap-induced shift of charge neutrality point (CNP) voltage $V_{CNP}$ [31]-[33], [41], [42] (for more details on the underlying model definitions see SI, §B).

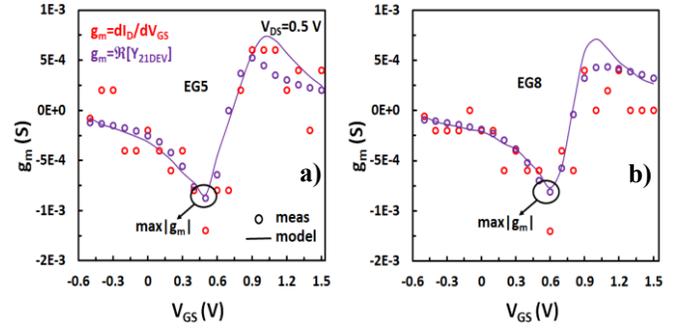

Fig. 2. Transconductance $g_m$ vs. $V_{GS}$ with markers representing the measurements (red: IV data, purple: Y parameters data at f=1 GHz) and lines the model for a) EG5 and b) EG8 GFETs for $V_{DS}$=0.5 V.

TABLE I
IV EXTRACTED PARAMETERS

| Parameter | Units | EG5 | EG8 |
|---|---|---|---|
| μ | cm$^2$/(V·s) | 200 | 170 |
| $C_{back}$ | μF/cm$^2$ | 1.87 | 1.87 |
| $V_{BS0}$ | V | 0.34 | 0.37 |
| $R_c$ | Ω | 134 | 134 |
| Δ | meV | 145 | 154 |
| hΩ | meV | 11 | 11 |
| Ktr | - | 0.35 | 0.5 |

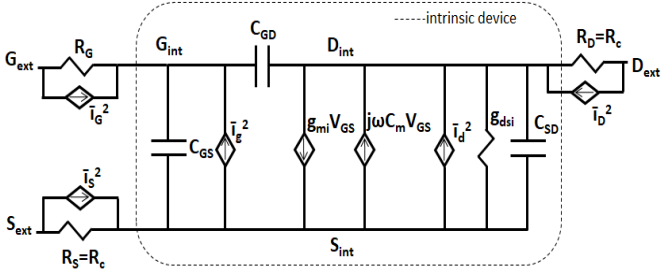

Fig. 3. Small signal Quasi-Static noise model for a GFET device. $g_{mi}$, $g_{dsi}$ are the intrinsic transconductance and output conductance respectively. $C_m = C_{DG} - C_{GD}$ where $C_{GS}$, $C_{GD}$, $C_{SD}$, $C_{DG}$ are the intrinsic capacitances [32]. Both extrinsic and intrinsic noise sources are shown. $<i_g^2>$, $<i_d^2>$, $<i_G^2>$, $<i_S^2>$, $<i_D^2>$: gate, drain, gate resistance $R_G$, source contact resistance $R_S$ and drain contact resistance $R_D$ current fluctuations, respectively.

### III. THERMAL CHANNEL NOISE MODEL

The basic procedure for the derivation of the total $S_{ID}$ is based on dividing the device channel into microscopic slices $\Delta x$, calculating all the local noise contributions at each $\Delta x$ and then integrating them along the gated channel region assuming a small-signal analysis since these local fluctuations are considered uncorrelated [19]-[21], [33], (see SI, §C1, Fig. S1b). The different noise sources of the GFET are presented in the small-signal circuit in Fig. 3 where apart from $S_{ID}$, channel induced gate current noise spectral density $S_{Ig}$ as well as the noise contributions from resistances $R_C$ and $R_G$ are included. Since the main contribution to minimum noise figure ($NF_{min}$), the measure of two port noise property, is stemming from $S_{ID}$ as $S_{ig}$ is negligible at f=1 GHz [16, Fig. 10-12], in this work we will enhance analysis of the spectral density of $I_D$ fluctuations. To calculate $S_{ID}$, a drift-diffusion current approach is used:

$$I_D = -W|Q_{gr}|\mu_{eff}E, \quad \mu_{eff} = \frac{\mu}{1+\frac{|E_x|}{E_c}}, \quad E_c = \frac{u_{sat}}{\mu} \quad (2)$$

where $|Q_{gr}|$ is the total graphene charge (see (A22) in SI, §B). The absolute value of $Q_{gr}$ indicates the movement of negative charged electrons and positive charged holes in opposite directions, additively contributing to the $I_D$. E, $E_x$, $E_c$ are the electric, the longitudinal electric and the critical electric fields respectively and $\mu_{eff}$ the effective mobility representing the degradation of the channel mobility at high electric field regime due to VS effect. The latter effect is considered in the proposed noise model since it is expected to increase $S_{ID}$ in short-channels at high $V_{DS}$ values [15]-[21]. A two-branch VS $u_{sat}$ model is used which is considered constant near CNP below a critical

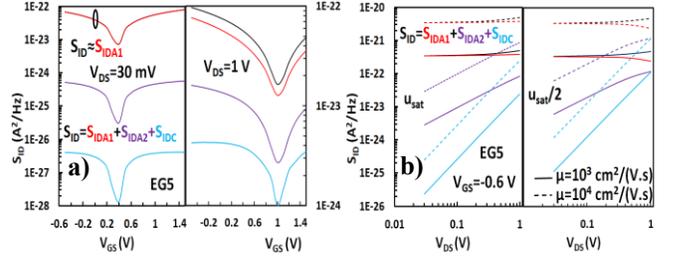

Fig. 4. Channel thermal noise $S_{ID}$ vs. a) $V_{GS}$ for low $V_{DS}$=30 mV (left subplot) and high $V_{DS}$=1 V (right subplot) and vs. b) $V_{DS}$ for $u_{sat}$ (left subplot) and $u_{sat}/2$ (right subplot) at two different increased mobility values, $\mu$=$10^3$, $10^4$ cm$^2$/(Vs), respectively for EG5 GFET at f=1 GHz. Total $S_{ID}$ and its different contributors are shown with different colors.

chemical potential value $V_{ccrit}$ and inversely proportional to chemical potential $V_c$ above $V_{ccrit}$ [33], (see (A24) in SI §B). Total $S_{ID}$ along the gated channel is given by [19, (6.3), (6.4)]:

$$S_{ID} = \int_0^L \frac{S_{\delta I_{nD}^2}(\omega,x)}{\Delta x}dx = \int_0^L G_{CH}^2 \Delta R^2 \frac{S_{\delta I_n^2}(\omega,x)}{\Delta x}dx \quad (3)$$

where $G_{CH}$ is the channel conductance and $\Delta R$ the resistance of the slice $\Delta x$ of the channel, $S_{\delta I_n^2}$ the PSD of the local noise source and $S_{\delta I_{nD}^2}$ the channel noise PSD due to a single noise source (see SI, §C1). $\mu_{eff}$ can be considered a function of both channel potential V and $I_D$ through $E_x(V, I_D)$ where the latter depends on the position x along the channel [19, §9.4.1], [20]. Thus, $I_D$ is defined as:

$$I_D = f(V, I_D) = \frac{W}{x}\int_{V_s}^V |Q_{gr}|\mu_{eff}(V, I_D)dV \Rightarrow dI_D = \frac{\partial f}{\partial V}dV + \frac{\partial f}{\partial I_D}dI_D \Leftrightarrow G_S = \frac{dI_D}{dV} = \frac{\partial f}{\partial V} + \frac{\partial f}{\partial I_D}\frac{dI_D}{dV} \Leftrightarrow G_S = \frac{W|Q_{gr}|\mu_{eff}}{x - W\int_{V_s}^V |Q_{gr}|\frac{\partial \mu_{eff}}{\partial I_D}dV} \quad (4)$$

since it can be easily shown from $I_D$ definition in (4) that:

$$\frac{\partial f}{\partial V} = \frac{W}{x}|Q_{gr}|\mu_{eff}, \quad \frac{\partial f}{\partial I_D} = \frac{W}{x}\int_{V_s}^V |Q_{gr}|\frac{\partial \mu_{eff}}{\partial I_D}dV \quad (5)$$

where $G_S$ is the transconductance on the source side (see SI, §C1, Fig. S1b). The next step would be to calculate $\partial \mu_{eff}/\partial I_D$ in the denominator of (4) [19], [20]:

$$\frac{\partial \mu_{eff}}{\partial I_D} = \frac{\partial \mu_{eff}}{\partial E_x}\frac{\partial E_x}{\partial E}\frac{\partial E}{\partial I_D} = \mu'_{eff}\frac{\partial E_x}{\partial E}\frac{\partial E}{\partial I_D} = \frac{\mu'_{eff}\frac{\partial E_x}{\partial E}}{\frac{\partial I_D}{\partial E}}, \quad \mu'_{eff} = \frac{\partial \mu_{eff}}{\partial E_x} \quad (6)$$

where from (2):

$$\frac{\partial I_D}{\partial E} = -W|Q_{gr}|\mu_{diff} \quad (7)$$

(for more details see (A29) in SI, §C2 with E. $\partial E_x/\partial E = E_x$ (see (A17) in SI, §B) where $\mu_{diff} = \mu_{eff} + \mu'_{eff}E_x$ [19, (9.3)] is the differential mobility. By using (6) and (7) in (4), $G_S$ yields:

$$G_S = \frac{W|Q_{gr}|\mu_{eff}}{x + \int_{V_s}^V \frac{\mu'_{eff}\frac{\partial E_x}{\partial E}}{\mu_{diff}}dV} \quad (8)$$

Similarly, the transconductance on the drain side $G_D$ [19], (see SI, §C1, Fig. S1b) can be calculated by following an identical procedure as in the $G_S$ case [19], [20]:

$$G_D = \frac{W|Q_{gr}|\mu_{eff}}{L - x + \int_V^{V_D}\frac{\mu'_{eff}\frac{\partial E_x}{\partial E}}{\mu_{diff}}dV} \quad (9)$$

and thus, $G_{CH}$ is given according to (8), (9):

$$\frac{1}{G_{CH}} = \frac{1}{G_S} + \frac{1}{G_D} \rightarrow G_{CH} = \frac{W|Q_{gr}|\mu_{eff}}{L + \int_{V_s}^{V_D}\frac{\mu'_{eff}\frac{\partial E_x}{\partial E}}{\mu_{diff}}dV} \quad (10)$$

(for more details see (A27) in SI, §C1, (A30) in SI, §C2). For $\Delta R$ calculation, (9) is applied from x to x+$\Delta x$ [20]. With the help of $\mu_{diff}$ definition and E. $\partial E_x/\partial E = E_x$ as mentioned before:

$$\Delta R = \frac{1}{\Delta G} = \frac{\Delta x}{W|Q_{gr}|\mu_{diff}} \quad (11)$$

(for more details see (A31) in SI, §C2).

In presence of an electric field, equilibrium does not stand anymore locally in the channel and thus, Einstein-relation between mobility and diffusion coefficient cannot be applied directly [19], [29], [30]. This can be dealt with the assumption of an Einstein-like expression to stand in nonequilibrium and the definition of a noise temperature $T_n \approx T_c$ where $T_c$ is the carrier temperature [19, (9.141, 9.142)]. For degenerate semiconductors like graphene, the contributions of total charge carriers to $I_D$ and $S_{\delta I_n^2}$ are no longer independent [28], [29] and thus, $S_{\delta I_n^2}$ must be multiplied with $\Delta \tilde{N}^2/\tilde{N}=(K_B T_L/n_{gr}) \cdot (\partial n_{gr}/\partial EF)$ where $\Delta \tilde{N}^2$ is the variance and $\tilde{N}$ the average number of carriers [29, (3)], [30]. $S_{\delta I_n^2}$ can be calculated if (11) is considered as [19, (6.13)]:

$$S_{\delta I_n^2}(\omega,x) = \frac{4K_B T_n}{\Delta R} = \frac{4K_B T_C}{\Delta R} \frac{\Delta \tilde{N}^2}{\tilde{N}} = 4K_B T_C \frac{W\mu_{diff} U_T}{\Delta x} k|V_c| \quad (12)$$

(for more details see (A32) in SI, §C2) where $T_L$ is the lattice (room) temperature, $n_{gr}=|Q_{gr}|/e$ [31] is the graphene charge density where $e$ is the elementary charge, $EF=e|V_c|$ is the shift of the Fermi level [32], (see SI, Fig. S1a), $U_T=K_B T_L/e$ the thermal voltage at room temperature and $k$ a coefficient [31], (see SI §B). Thus, (3) is transformed because of (10)-(12) as:

$$S_{ID} = 4K_B T_C U_T k \frac{W}{L^2} \int_0^L \frac{\mu_{eff}^2}{\mu_{diff}} M |V_c| \, dx \quad (13)$$

where $\mu^2_{eff}/\mu_{diff}=\mu$ [41, (A36) in SI, §C2]. $M$ is given by [19]:

$$M = \frac{1}{\left(1+\frac{1}{L}\int_{V_S}^{V_D}\frac{\mu'_{eff}}{\mu_{diff}}\frac{\partial E_x}{\partial E}dV\right)^2} = \frac{1}{\left(1+\frac{\mu}{CL}\int_{V_{cs}}^{V_{cd}}\frac{C_q}{u_{sat}}|dV_c|\right)^2} = \left(\frac{L}{L_{eff}}\right)^2 \quad (14)$$

(see (A37), (A38) in SI, §C2) where $C_q=k|V_c|$ is the quantum capacitance [31]-[33], (see SI, §B, Fig. S1a), C is the sum of top and back oxide capacitances, $L_{eff}$ accounts for an effective channel length representing the reduction of $I_D$ due to VS effect [31]-[33] and thus, two cases shall be considered for its solution according to the two-branch $u_{sat}$ model applied in this work (see (A24), (A25) in SI, §B). $V_{cs}$, $V_{cd}$ are the chemical potentials at source and drain sides, respectively (see, (A23) in SI, §B). From [19, (9.150)], (see (A17) in SI, §B):

$$\mu_{eff} = \mu\sqrt{\frac{T_L}{T_C}} \Rightarrow \frac{T_C}{T_L} = \left(\frac{\mu}{\mu_{eff}}\right)^2 = \left(1+\frac{|E_x|}{E_C}\right)^2 \quad (15)$$

and (13) becomes due to (14), (15):

$$S_{ID} = 4K_B T_L U_T k\mu \frac{W}{L_{eff}^2}\int_0^L \left(1+\frac{|E_x|}{E_C}\right)^2 |V_c| dx =$$

$$4K_B T_L U_T k\mu \frac{W}{L_{eff}^2}\left[\int_0^L |V_c|dx + \int_0^L 2\frac{|E_x|}{E_C}|V_c|dx + \int_0^L \left(\frac{E_x}{E_C}\right)^2 |V_c|dx\right] \quad (16)$$

Integral in (16) can be split into three integrals named $S_{IDA}$, $S_{IDB}$, $S_{IDC}$. In order to solve each one of them, the integral variable change from x to $V_c$ shall be applied (see (A19) in SI, §B). Thus:

$$S_{IDA} = 4K_B T_L U_T k\mu \frac{W}{L_{eff}^2}\int_0^L |V_c|dx =$$

$$4K_B T_L U_T k\mu \frac{W}{L_{eff}^2}\left[\int_{V_{cs}}^{V_{cd}}\left(-\frac{|V_c||Q_{gr}|2L_{eff}}{kg_{vc}}\left(\frac{C_q+C}{C}\right)\right)dV_C - \int_{V_{cs}}^{V_{cd}}\left(\frac{\mu|V_c|}{u_{sat}}\left(\frac{C_q}{C}\right)\right)|dV_c|\right] \quad (17)$$

which again is split into two integrals, namely $S_{IDA1}$ (1st in the brackets) and $S_{IDA2}$ (2nd in the brackets) as $S_{IDA}= S_{IDA1}- S_{IDA2}$ where:

$$S_{IDA1} = 4K_B T_L U_T k\mu \frac{W}{Cg_{vc}L_{eff}}\int_{V_{cd}}^{V_{cs}}\left(|V_c|(k|V_c|+C)\left(V_c^2+\frac{\alpha}{k}\right)\right)dV_C = 4K_B T_L U_T k\mu \frac{W}{Cg_{vc}L_{eff}}\left[\pm\frac{\alpha C V_c^2}{2k}+\frac{\alpha V_c^3}{3}\pm\frac{CV_c^4}{4}+\frac{kV_c^5}{5}\right]_{V_{cd}}^{V_{cs}} \quad (18)$$

where $gV_c$ is a normalized $I_D$ term (see (A21) in SI, §B) and $\alpha$ is related to residual charge [31]-[33]. VS effect contributes to $S_{IDA1}$ only through $L_{eff}$ while for $S_{IDA2}$ both $L_{eff}$ and $u_{sat}$ are included. As in $L_{eff}$ solution (see (A25) in SI, §B), two cases shall be considered for the solution of $S_{IDA2}$ according to the two-branch $u_{sat}$ model (see (A24) in SI, §B). Thus, near CNP

$$S_{IDA2} = 4K_B T_L U_T k\mu^2 \frac{W}{CL_{eff}^2}\int_{V_{cs}}^{V_{cd}}\left(\frac{kV_c^2}{S}\right)|dV_c| = 4K_B T_L U_T k\mu^2 \frac{W}{CSL_{eff}^2}\left|\left[\frac{kV_c^3}{3}\right]_{V_{cd}}^{V_{cs}}\right| \to |V_c|<V_{ccrit} \quad (19a)$$

whereas away CNP:

$$S_{IDA2} = 4K_B T_L U_T k\mu^2 \frac{W}{CL_{eff}^2}\int_{V_{cs}}^{V_{cd}}\left(\frac{kV_c^2\sqrt{V_c^2+\frac{\alpha}{k}}}{N}\right)|dV_c| =$$

$$4K_B T_L U_T k\mu^2 \frac{W}{CNL_{eff}^2}\left|\left[\frac{1}{8k}\left(kV_c\sqrt{V_c^2+\alpha/k}(\alpha+2kV_c^2)-\alpha^2\ln(V_c+\sqrt{V_c^2+\alpha/k})\right)\right]_{V_{cd}}^{V_{cs}}\right| \to |V_c|>V_{ccrit} \quad (19b)$$

(for S, N definition see (A24), in SI, §B). The absolute value in the analytical solution of (19) comes from $|dV_c|$ in order to distinguish two cases for $S_{IDA2}$ depending on the sign of $dV_c$. Thus, in the case of $dV_c<0 \to V_{cs}>V_{cd}$ ($V_{DS}>0$) $\to |dV_c|=-dV_c$, the integral is solved from $V_{cd}$ to $V_{cs}$ while when $dV_c>0 \to V_{cs}<V_{cd}$ ($V_{DS}<0$) $\to |dV_c|=dV_c$, the integral is solved from $V_{cs}$ to $V_{cd}$. For the solution of $S_{IDC}$ (see (A40) in SI, §C3), the main idea was to express electric field as: $E^2=(-dV/dx)(-dV/dx)$ (see (A17) in SI, §B) and then both sides are integrated after being multiplied with dx which has as a result a double integral notation. $S_{IDC}$ is directly affected by the square of $u_{sat}$, thus again two different cases shall be considered. Near CNP:

$$S_{IDC} = 4K_B T_L U_T k\mu^3 \frac{W}{LC^2 L_{eff}^2}\int_{V_{cs}}^{V_{cd}}\int_{V_{cs}}^{V_{cd}}\frac{(k|V_c|)^3}{S^2}|dV_c||dV_c| =$$

$$4K_B T_L U_T k\mu^3 \frac{W}{S^2 LC^2 L_{eff}^2}(V_{cs}-V_{cd})\left[\pm\frac{k^2 V_c^4}{4}\right]_{V_{cd}}^{V_{cs}} \to |V_c|<V_{ccrit} \quad (20a)$$

and away CNP:

$$S_{IDC} = 4K_B T_L U_T k\mu^3 \frac{W}{LC^2 L_{eff}^2}\int_{V_{cs}}^{V_{cd}}\int_{V_{cs}}^{V_{cd}}\frac{(k|V_c|)^3(V_c^2+\alpha/k)}{N^2}|dV_c||dV_c| =$$

$$4K_B T_L U_T k\mu^3 \frac{W}{N^2 LC^2 L_{eff}^2}(V_{cs}-V_{cd})\left[\pm k\left(\frac{\alpha V_c^4}{4}+\frac{kV_c^6}{6}\right)\right]_{V_{cd}}^{V_{cs}} \to |V_c|>V_{ccrit} \quad (20b)$$

Oppositely with $S_{IDA2}$, $S_{IDC}$ has always the same solution regardless of $V_{DS}$ polarity, since the sign of the product $|dV_c||dV_c|=dV_c.dV_c$ for $dV_c>0$ ($V_{DS}<0$) or $|dV_c||dV_c|=(-dV_c)(-dV_c)$ for $dV_c<0$ ($V_{DS}>0$) is always positive. In $\pm, \mp$ notation in (18)-(20), top sign refers to $V_c>0$ and bottom sign to $V_c<0$ case. It can be easily shown that $S_B=2S_{A2}$ (see (A39) in SI, §C3) which means that $S_{ID}= S_{IDA}+ S_{IDB}+ S_{IDC}= S_{IDA1}+ S_{IDA2}+ S_{IDC}$.

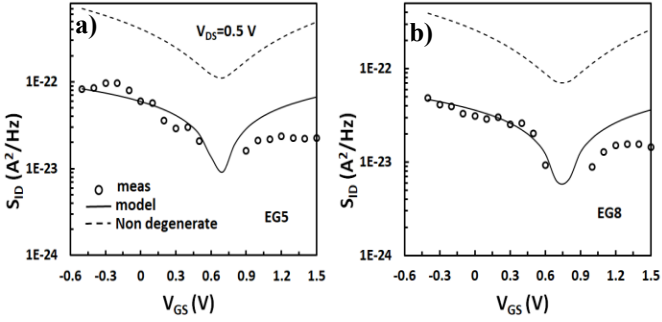

Fig. 5. $S_{ID}$ vs. $V_{GS}$ for a) EG5 and b) EG8 GFETs for $V_{DS}$=0.5 V and f=1 GHz. markers: measured, solid lines: model, dashed lines: Non-degenerate model.

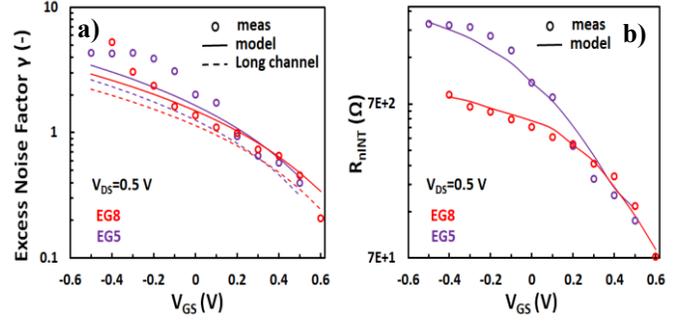

Fig. 6. Noise excess factor γ (a) and intrinsic noise resistance $R_{nINT}$ (b) vs. $V_{GS}$ for EG5 (red) and EG8 (purple) GFETs for $V_{DS}$=0.5 V and f=1 GHz. markers: measured, solid lines: model, dashed lines: long-channel model.

$S_{IDA1}$, $S_{IDA2}$ and $S_{IDC}$ in (18)-(20), respectively can be solved in a compact way which is of outmost importance for circuit designers. It is critical to mention that the signs and the absolute values of $V_{cs}$, $V_{cd}$ define different regions where $V_{cs}$, $V_{cd}$ belong (p- or n-type, above or below $V_{ccrit}$) and the integrals shall be solved in each of these regions and then added (see SI, §E of [33]). $S_{ID}$ as well as its contributors ($S_{IDA1}$, $S_{IDA2}$ and $S_{IDC}$) are illustrated in Fig. 4a for both low and high $V_{DS}$ at left and right subplots, respectively vs. $V_{GS}$ for the EG5 GFET where the simulations were conducted with the IV parameters from Table I. It is apparent that in low electric field regime, VS induced terms $S_{IDA2}$ and $S_{IDC}$ are negligible and $S_{ID} \approx S_{IDA1}$. $S_{IDA1}$ has a dependence on VS through $L_{eff}$ but in low $V_{DS}$, $L \approx L_{eff}$. At $V_{DS}$=1 V, $S_{ID2}$ and $S_{IDC}$ have become significant and they increase total $S_{ID}$. It is also clear that each noise term increases as we go deeper in p- and n-type region, respectively. In addition to the model validation with the low mobility DUTs used in this work, cf. Table I, we have benchmarked our model at two higher μ ($10^3$, $10^4$ cm²/(V.s)) values. HFN related terms are shown in Fig. 4b vs. $V_{DS}$ at $V_{GS}$=-0.6 V where $S_{ID}$ is maximum, cf. Fig. 4a, for EG5 GFET. All noise contributors increase with μ as it is predicted from (16) and shown in Fig. 4b while any additive increase of $(1+|E_x|/E_c)^2$ term with μ in numerator of (16) through $E_c$, cf. (2), is largely counterbalanced from the corresponding increment of $L^2_{eff}$ in the denominator of (16). In the right subplot of Fig. 4b, a $u_{sat}/2$ case is shown, where VS effect is more acute than the $u_{sat}$ case (left subplot) mainly due to a steeper $S_{IDA1}$ reduction with $V_{DS}$ caused by the $\sim 1/L_{eff} \rightarrow \sim u_{sat}$ trend of $S_{IDA1}$, cf. (18). $S_{IDA2}$, $S_{IDC}$ exhibit a direct $\sim 1/u_{sat}$ and $\sim 1/u^2_{sat}$ dependence respectively, through S, N VS-related parameters while the concurrent $\sim 1/L^2_{eff} \rightarrow \sim u^2_{sat}$ contribution, cf. (19)-(20), leads to a $\sim u_{sat}$ trend and thus, to a saturation of $S_{IDA2}$ at high $V_{DS}$, and to no decrease for $S_{IDC}$ with $V_{DS}$ as the different effects are compensated there.

## IV. RESULTS – DISCUSSION

The proposed $S_{ID}$ model is validated with experimental data from two short-channel GFETs in this section. As mentioned before, the measurement frequency f=1 GHz primarily ensures the QS region of operation which results in a frequency independent behavior of $S_{ID}$. At Non-QS regime, one should deal with induced gate noise as well as carrier inertia effects which would produce different current noise PSDs at source and drain; the latter is not the purpose of the present study.

Thus, in the following plots the attention is focused on the bias dependence of noise. In Fig. 5, both measured and simulated $S_{ID}$ are depicted vs. $V_{GS}$ at $V_{DS}$=0.5 V from strong p-type to strong n- type regime for both devices where asymmetries of IV [33] and Y-parameters data, and consequently $S_{ID}$ data are recorded. Additionally, the reduced gain due to low |$g_m$| near CNP does not ensure accurate $R_n$ measurements there since a sufficient gain is required for the Y-factor HFN measurement method, thus $V_{GS}$ points very close to CNP ($V_{GS}$=0.6, 0.7 V for EG5 and $V_{GS}$=0.7, 0.8 V for EG8) are omitted. Experimental data are extracted from (1) at any bias point since noise and Y-parameters are measured simultaneously from the same set-up while simulated data are obtained by solving (16) with (18)-(20). The model provides accurate description of the experiments for both GFETs in p-type regime where we focus our analysis since maximum $g_m$ is estimated there, cf. Fig. 2. This work for the first time considers the degenerate nature of graphene and how it affects $S_{ID}$ performance thus, the non-degenerate case is also shown in Fig. 5 for comparison reasons. For more details on the extraction of the non-degenerate $S_{ID}$ model see SI, §C4 where the contributions of total charge carriers to $I_D$ and $S_{δI^2_n}$ are independent, thus $S_{δI^2_n}$ is not multiplied with $\Delta \tilde{N}^2/\tilde{N}$ [19], [29], [30]. Non-degenerate case (dashed lines in the Fig. 5) overestimates $S_{ID}$ almost one order of magnitude for both devices. These results clearly show that the application of $S_{ID}$ models taken from MOSFETs related noise models with assumption of non-degenerate channel directly to GFETs is not valid.

An important Figure of Merit (FoM) for RF circuit design for noise performance is an excess noise factor γ, introduced by *Van der Ziel* [12] and widely investigated in CMOS devices [13], [19]-[21], [39]:

$$\gamma = \frac{g_n}{g_{mi}}, g_n = \frac{S_{ID}}{4KT_L} \qquad (21)$$

where $g_{mi}$ is the intrinsic transconductance of the device (removed $R_c$ and $R_G$ resistances) and $g_n$ is noise conductance [19]-[21]. The latter is defined by the $S_{ID}$ in (16) with (18)-(20), divided by $4K_BT_L$. Initially, excess noise factor was referred as α whereas γ was the thermal noise parameter defined as $g_n/g_{dso}$ where $g_{dso}$ is the output conductance at $V_{DS}$=0 V [12]-[18]. Thermal noise parameter is not an ideal FoM for analog/RF design since $g_n$ and $g_{dso}$ are evaluated at different operating conditions [19], [21]. Excess noise factor is of outmost importance for noise performance in RF circuits since it accounts for the generated noise at the drain side of the device for a given transconductance [19], [21], [39]. $g_n$ and $g_{mi}$ can be evaluated for the same bias point and $\gamma$ is important to determine the noise figure (NF) of an LNA [21]. In this work, excess noise factor $\gamma$ is for the first time characterized for GFETs. Consequently, measured and modelled $\gamma$ are displayed in Fig.6a vs. $V_{GS}$ at $V_{DS}$=0.5 V for DUTs. Data are extracted by using (1) for $S_{ID}$ in (21) while model by using (16) with (18)-(20) in (21). Measured $\gamma$ can be up to ~3 to 4 showing an increasing trend with higher carrier densities. The model precisely captures this behaviour. Simulated $\gamma$ for long-channel case is presented with dashed lines by de-activating VS effect (hΩ parameter) in our model, cf. Fig. 6a. This leads to an underestimation of $\gamma$ by up to 30%, compared to measured data. Noise resistance behavior versus bias is depicted in Fig. 6b. Taking into account (1) and (21), $R_{nINT}$ can be calculated as:

$$R_{nINT} = \frac{g_n}{|Y21_{INT}|^2} \quad (22)$$

and then compared to the measured $R_{nINT}$ cf. Fig. 6c-6d. The results present a consistency of the model vs. measured data, whereas $R_{nINT}$ increases towards to stronger p-type region.

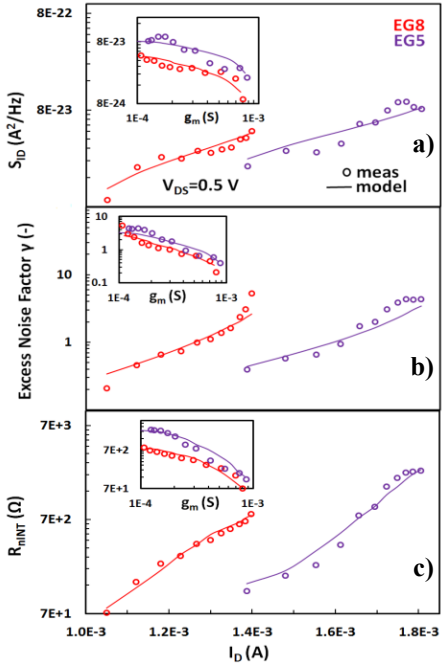

Fig. 7. $S_{ID}$ (a), $\gamma$ (b) and $R_{nint}$ (c) vs. drain current $I_D$ and vs. $g_m$ in insets for EG5 (red) and EG8 (purple) GFETs for $V_{DS}$=0.5 V and f=1 GHz. markers: measured, solid lines: model.

For more explicit analysis, $S_{ID}$, $\gamma$ and $R_{nINT}$ for both investigated DUTs are shown vs. $I_D$ and $g_m$ (insets) in Fig. 7a, 7b and 7c respectively for $V_{DS}$=0.5 V. The proposed GFET noise model accounts well the measured data: $S_{ID}$, $\gamma$ and $R_{nINT}$ dependences on $I_D$, cf. Fig. 7. Presented parameters increase with $I_D$ and such trend agrees with results from MOSFETs [15]-[21]. In terms of $S_{ID}$, there is a saturation-like trend at higher $I_D$ values (or at strong p-type region as shown in Fig. 5) which also agrees with findings from CMOS [15]-[18], [21]. Moreover, the shortest device (EG5) exhibits higher noise as it was expected. This is more evident in the insets of Fig. 7 vs. $g_m$ since both GFETs appear to have similar $g_m$ while maximum $g_m$ value corresponds to minimum noise.

## V. CONCLUSIONS

A complete physics-based analytical intrinsic channel thermal noise model in QS region of operation for GFETs is derived and verified on the measured data. The presented $S_{ID}$ model describes precisely the bias dependence of noise, including VS effect while it also considers the degenerate nature of graphene for the first time. The proposed model can be easily implemented in Verilog-A for the use with circuit simulators. The model is successfully validated with experimental high frequency Y-parameters and noise data without the need of any fitting parameter which proves its physical consistency. Noise PSD increases with $I_D$ and saturates at deep p-type region similarly to CMOS devices. Apart from $S_{ID}$, noise excess factor $\gamma$ is defined for the first time for GFETs. Its value for the short-channel GFETs under test reaches maximum from ~3 to 4 at higher $I_D$ (EG5: ~1.8 mA, EG8: ~1.4 mA) away from CNP whereas it is lower for smaller currents near CNP (EG5: ~1.4 mA, EG8: ~1.1 mA). This trend is successfully predicted by the model whereas the simulations without taking into account VS effect reveal an underestimation of $\gamma$ around 30%. Furthermore, $S_{ID}$, $\gamma$ and $R_{nINT}$ present a minimum at highest $g_m$ value which is a very useful information for the circuit design point of view. These quantities are higher for the shortest device and the proposed model fits accurately this characteristic which is indicative of a proper scaling behaviour. GFET HFN studied in this work, shows comparable results with CMOS [20] indicating that this emerging technology is on a good track of development and could eventually compete with incumbent devices without facing the scaling limitations of the latter.

ACKNOWLEDGMENT

This work was funded by the European Union's Horizon 2020 research and innovation program under Grant Agreement No. GrapheneCore2 785219 and No. GrapheneCore3 881603. We also acknowledge financial support by Spanish government under the projects RTI2018-097876-B-C21 (MCIU/AEI/FEDER, UE) and project 001-P-001702-GraphCat: Comunitat Emergent de Grafè a Catalunya, co-funded by FEDER within the framework of Programa Operatiu FEDER de Catalunya 2014-2020. This work was partly supported by the French RENATECH network.

## Supplementary Information for:

## Bias-dependent intrinsic RF thermal noise modeling and characterization of single layer graphene FETs

*Nikolaos Mavredakis[1], Anibal Pacheco-Sanchez[1], Paulius Sakalas[2-4], Wei Wei[5], Emiliano Pallecchi[5], Henri Happy[5] and David Jiménez[1]

[1] Departament d'Enginyeria Electrònica, Escola d'Enginyeria, Universitat Autònoma de Barcelona, Bellaterra 08193, Spain
[2] MPI AST Division, 302 Dresden, Germany
[3] Semiconductor Physics Institute of Center for Physical Sciences and Technology, LT-10257, Vilnius, Lithuania
[4] Baltic Institute for Advanced Technologies, LT-01403, Vilnius, Lithuania
[5] Institute of electronics, Microelectronics and Nanotechnology, CNRS UMR8520, 59652 Villeneuve d'Ascq, France
E-mail: nikolaos.mavredakis@uab.es

### A. Supplementary Information: De-embedding process and extraction of intrinsic $|Y21_{INT}|$ and noise resistance $R_{nINT}$ parameters

### A1. RF and noise de-embedding

De-embedding is the process of removing unwanted parasitics from high frequency measurements and has to be applied at the Device Under Test (DUT) before the extraction of any device parameter. After RF and noise de-embedding, the de-embedded parameters will be referred as DEV parameters. In this study, an open-short-pad de-embedding procedure was applied for the extraction of $Y_{DEV}$ parameters [35]-[36] since OPEN, SHORT and PAD structures and data were available for the DUT. In more detail, measured S-parameters ($S_{DUT}$) are transformed to Y-parameters ($Y_{DUT}$) and then the following equation is applied:

$$Y_{DEV} = [(Y_{DUT} - Y_{PAD})^{-1} - (Y_{SHORT} - Y_{PAD})^{-1}]^{-1} - [(Y_{OPEN} - Y_{PAD})^{-1} - (Y_{SHORT} - Y_{PAD})^{-1}]^{-1}$$

(A1)

It is crucial to mention here that for noise de-embedding, S-parameters have to be measured together with noise parameters (HFN) since $Y_{DEV}$ parameters participate in the noise de-embedding, as it will be shown later. For noise de-embedding, an open methodology was applied which is based on the noise correlation matrix approach [37] according to the following steps [38]:

1. Calculation of the measured correlation matrix $C_{ADUT}$

$$C_{ADUT} = 2K_B T \begin{bmatrix} R_{nDUT} & \frac{NFmin_{DUT}-1}{2} - R_{nDUT}(Y_{optDUT})^* \\ \frac{NFmin_{DUT}-1}{2} - R_{nDUT}Y_{optDUT} & R_{nDUT}|Y_{optDUT}|^2 \end{bmatrix}$$

(A2)

where $R_{nDUT}$, $N_{FminDUT}$, $Y_{OPTDUT}$ are the measured HFN parameters while * symbolizes the complex conjugate.

2. Conversion of $C_{ADUT}$ to $C_{YDUT}$:

$$C_{YDUT} = T_{DUT}C_{ADUT}T_{DUT}{}^\dagger, \quad T_{DUT} = \begin{bmatrix} -Y_{11DUT} & 1 \\ -Y_{21DUT} & 0 \end{bmatrix} \tag{A3}$$

where † corresponds to the transpose and complex conjugate matrix.

3. Calculation of $C_{YOPEN}$:

$$C_{YOPEN} = 2K_B T \mathcal{R}(Y_{OPEN}) \tag{A4}$$

4. De-embedding of $C_{YDEV}$ as:

$$C_{YDEV} = C_{YDUT} - C_{YOPEN} \tag{A5}$$

5. Conversion of $Y_{DEV}$ from (A1) to chain matrix A as:

$$A_{DEV} = \frac{-1}{Y_{21DEV}} \begin{bmatrix} Y_{22DEV} & 1 \\ Y_{11DEV}Y_{22DEV} - Y_{12DEV}Y_{21DEV} & Y_{11DEV} \end{bmatrix} \tag{A6}$$

6. Conversion of $C_{YDEV}$ to $C_{ADEV}$

$$C_{ADEV} = T_{ADEV}C_{YDEV}T_{ADEV}{}^\dagger, \quad T_{ADEV} = \begin{bmatrix} 0 & A_{12DEV} \\ 1 & A_{22DEV} \end{bmatrix} \tag{A7}$$

7. Calculation of noise de-embedded parameters $HFN_{DEV}$ from $C_{ADEV}$

$$C_{ADEV} = 2K_B T \begin{bmatrix} R_{nDEV} & \frac{NF_{minDEV}-1}{2} - R_{nDEV}(Y_{optDEV})^* \\ \frac{NF_{minDEV}-1}{2} - R_{nDEV}Y_{optDEV} & R_{nDEV}|Y_{optDEV}|^2 \end{bmatrix} \tag{A8}$$

It is apparent from (A6)-(A8) that de-embedded $Y_{DEV}$ parameters participate directly into HFN parameters de-embedding.

**A2. Contact and gate resistance removal from de-embedded Y and noise resistance parameters**

Whereas in CMOS technologies, contact resistance $R_c$ effect on $Y_{DEV}$ parameters can be neglected since $R_c$ is very low, this is not the case for GFETs thus, this effect shall be removed. Regarding $Y_{DEV}$, the procedure proposed in Ref. [36, (10)-(13)] is applied to remove $R_c$ effect. After removing $R_c$ contribution from $Y_{DEV}$, intrinsic Y-parameters $Y_{INT}$ can be calculated as:

$$Y_{INT(RG)} =$$
$$\begin{bmatrix} \omega^2 R_G C_{GG}^2 + j\omega C_{GG} & -\omega^2 R_G C_{GG}C_{GD} - j\omega C_{GD} \\ g_{mi} - \omega^2 R_G C_{GG}C_{DG} - j\omega(C_{DG} + g_{mi}R_G C_{GG}) & g_{dsi} + \omega^2 R_G C_{GG}(C_{GD} + C_{SD}) + j\omega(C_{GD} + C_{SD} - g_{dsi}R_G C_{GG}) \end{bmatrix}$$
(A9)

As it can be observed from (A9), gate resistance $R_G$ contribution to $Y_{INT}$ parameters is still there and has also to be removed. Intrinsic Y-parameters $Y_{INT}$ without $R_G$ contribution are given by [32]:

$$Y_{INT} = \begin{bmatrix} j\omega C_{GG} & -j\omega C_{GD} \\ g_{mi} - j\omega C_{DG} & g_{dsi} + j\omega(C_{GD} + C_{SD}) \end{bmatrix} \tag{A10}$$

Imaginary parts of $Y_{11INT(RG)}$, $Y_{12INT(RG)}$ and $Y_{11INT}$, $Y_{12INT}$ in (A9-A10), respectively are the same thus $C_{GG}$, $C_{GD}$ and $R_G$ can be extracted:

$$C_{GG} = \frac{\Im(Y_{11INT})}{\omega}, \; C_{GD} = \frac{-\Im(Y_{12INT})}{\omega}, \; R_G = \frac{\Re(Y_{11INT(RG)})}{\Im(Y_{11INT})^2} \tag{A11}$$

$C_{DG}$ can be calculated by imaginary part of $Y_{21INT(RG)}$ in (A9) as:

$$\Im(Y_{21INT(RG)}) = -\omega(C_{DG} + g_{mi}R_G C_{GG}) \rightarrow C_{DG} = -\frac{\Im(Y_{21INT(RG)})}{\omega} - g_{mi}R_G C_{GG} \tag{A12}$$

whereas $g_{mi}$ from the real part of $Y_{21INT(RG)}$ in (A9) if (A12) is used:

$$\Re(Y_{21INT(RG)}) = g_{mi} - \omega^2 R_G C_{GG} C_{DG} = g_{mi} - \omega^2 R_G C_{GG}\left(-\frac{\Im(Y_{21INT(RG)})}{\omega} - g_{mi}R_G C_{GG}\right) \tag{A13}$$

In (A13), $g_{mi}$ is calculated as it is the only unknown term and then from (A12) $C_{DG}$ is also extracted. Thus, $Y_{21INT}$ in (A10) is calculated which is essential for the intrinsic channel noise, as shown in (1) of the main manuscript. For the complete characterization of $Y_{INT}$ parameters in (A10), $Y_{22INT}$ must be extracted. Thus, $C_{GD} + C_{SD}$, and consequently $C_{SD}$, can be calculated by imaginary part of $Y_{22INT(RG)}$ in (A9) as:

$$\Im(Y_{22INT(RG)}) = \omega(C_{GD} + C_{SD} - g_{dsi}R_G C_{GG}) \rightarrow C_{GD} + C_{SD} = \frac{\Im(Y_{22INT(RG)})}{\omega} + g_{dsi}R_G C_{GG} \tag{A14}$$

whereas $g_{dsi}$ from the real part of $Y_{22INT(RG)}$ in (A9) if (A14) is used:

$$\Re(Y22_{INT(RG)}) = g_{dsi} + \omega^2 R_G C_{GG}(C_{GD} + C_{SD}) = g_{dsi} + \omega^2 R_G C_{GG}\left(\frac{\Im(Y22_{INT(RG)})}{\omega} + g_{dsi}R_G C_{GG}\right) \tag{A15}$$

In (A15), $g_{dsi}$ is calculated as it is the only unknown term and then from (A14) $C_{SD}$ is also extracted.

For the purposes of this study, only intrinsic noise resistance parameter $R_{nINT}$ is needed for the derivation of intrinsic channel thermal noise as it is shown in (1) of the main manuscript. Intrinsic noise resistance is calculated as if (A8) is considered [40]:

$$R_{nINT} = R_{nDEV} - R_C - R_G = \frac{C_{A11DEV}}{2K_B T} - R_C - R_G \tag{A16}$$

**B. Supplementary Information: Definitions of basic quantities of the IV model**

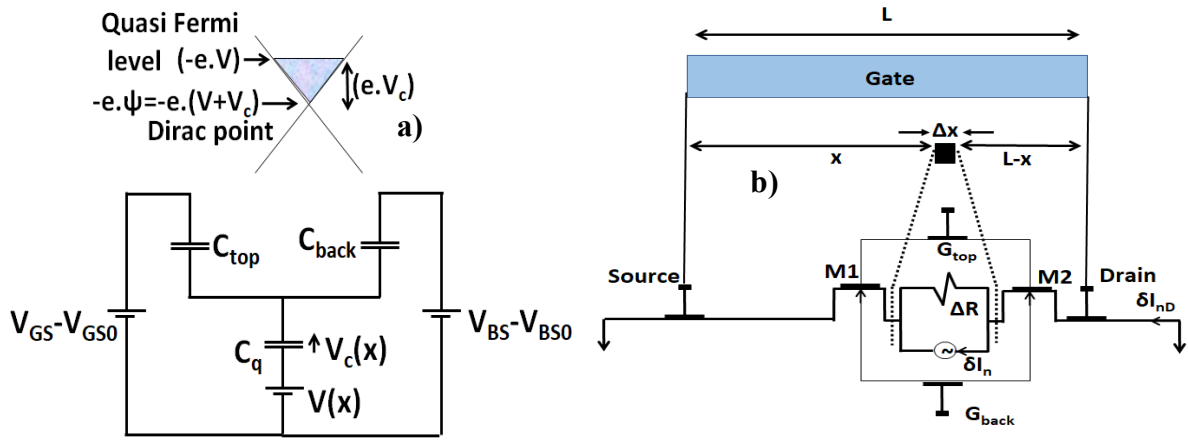

**Fig. S1.** a) Energy dispersion diagram of GFET (top) and its capacitive circuit (bottom) are shown. b) The equivalent circuit for a local current noise contribution to the total noise is illustrated. Each noise-generating slice of the channel is connected to two noiseless GFETs, M1 and M2 respectively.

Fig. S1a depicts the equivalent capacitive circuit of the CV-IV chemical potential-based model [31], [32] where quantum capacitance ($C_q$) is the derivative of graphene net charge $Q_{net}$ and chemical potential $V_c(x)$.

A linear relationship is considered between $C_q$ and $V_c$ ($C_q=k|V_c|$) where $k=2e^3/(\pi h^2 u_f^2)$ [31] with e the elementary charge, $u_f$ the Fermi velocity ($=10^6$ m/s) and h the reduced Planck constant ($=1.05 \cdot 10^{-34}$ J.s). $V_c(x)$ corresponds to the voltage drop across $C_q$ at channel position x and equals to the potential difference between the quasi-Fermi level and the potential at the CNP, as illustrated in the energy dispersion relation scheme of graphene in the top drawing of Fig. S1a where $V_c(0)=V_{cs}$ and $V_c(L)=V_{cd}$ at the Source (x=0) and Drain (x=L) end, respectively. Top and back gate source voltage overdrives are represented as: $V_{GS}-V_{GS0}$, $V_{BS}-V_{BS0}$ whereas top and back gate capacitances as: $C_{top}$ and $C_{back}$ where $C=C_{top}+C_{back}$. $V(x)$ is the graphene channel quasi-Fermi potential at position x, which equals to zero at the Source and $V_{DS}$ at the Drain end respectively. It is known that [31]-[33]:

$$E = -\frac{dV}{dx}, \quad E_c = \frac{u_{sat}}{\mu}, \quad E_x = -\frac{d\psi}{dx}, \quad \frac{dV}{dV_c} = -\frac{C_q+C}{C} \rightarrow \frac{dV_c}{dV} = -\frac{C}{C_q+C}, \quad \frac{d\psi}{dV_c} = \frac{dV+dV_c}{dV_c} = -\frac{C_q}{C}, \quad \frac{E_x}{E} = \frac{-\frac{d\psi}{dx}}{-\frac{dV}{dx}} =$$

$$\frac{d\psi}{dV} = \frac{d\psi}{dV_c}\frac{dV_c}{dV} = \frac{C_q}{C_q+C} \rightarrow \frac{dE_x}{dE}E = E_x \quad (A17)$$

where $\psi=V+V_c$ is the electrostatic potential and all other quantities are defined in the main manuscript. Equation (2) of the main manuscript is transformed due to (A17):

$$I_D = -W|Q_{gr}|\mu_{eff}E = W|Q_{gr}|\mu_{eff}\frac{dV}{dx} = W|Q_{gr}|\frac{\mu}{1+\frac{\mu}{v_{sat}}\left|-\frac{d\psi}{dx}\right|}\frac{dV}{dx} \leftrightarrow I_D\left[1+\frac{\mu}{v_{sat}}\left|-\frac{d\psi}{dV_c}\frac{dV_c}{dx}\right|\right] =$$

$$W|Q_{gr}|\mu\frac{dV}{dV_c}\frac{dV_c}{dx} \leftrightarrow I_D\left[1+\frac{\mu}{v_{sat}}\frac{C_q}{C}\left|\frac{dV_c}{dx}\right|\right] = -W|Q_{gr}|\mu\frac{C_q+C}{C}\frac{dV_c}{dx} \text{ where } I_D = \frac{\mu Wkg_{vc}}{2L_{eff}} \text{ [30]} \quad (A18)$$

Then from (A17), (A18) we can end up with equation:

$$\frac{dx}{dV_c} = \frac{-|Q_{gr}|2L_{eff}}{kg_{vc}}\left(\frac{C_q+C}{C}\right) - \frac{\mu}{v_{sat}}\left(\frac{C_q}{C}\right)\left|\frac{dV_c}{dV_c}\right| \quad (A19)$$

If we multiply both terms of (A18) with dx and then integrate from Source to Drain we get:

$$I_D = \frac{-W\mu\int_{V_{cs}}^{V_{cd}}|Q_{gr}|\frac{C_q+C}{C}dV_c}{L+\mu\int_{V_{cs}}^{V_{cd}}\frac{C_q}{v_{sat}C}|dV_c|} = \frac{W\mu\int_{V_{cd}}^{V_{cs}}|Q_{gr}|\frac{k|V_c|+C}{C}dV_c}{L+\mu\int_{V_{cs}}^{V_{cd}}\frac{k|V_c|}{u_{sat}C}|dV_c|} \quad (A20)$$

Bias dependent term $gV_c$ in (A18), (A19) which expresses the normalized drain current, is in fact the numerator integral in (A20) and is calculated as [31]:

$$gV_c = [g(V_c)]_{V_{cd}}^{V_{cs}} + \frac{\alpha V_{DS}}{k} = \frac{V_{cs}^3-V_{cd}^3}{3} + \frac{k}{4C}\left[\text{sgn}(V_{cs})V_{cs}^4 - \text{sgn}(V_{cd})V_{cd}^4\right] + \frac{\alpha V_{DS}}{k} \quad (A21)$$

Graphene charge is given by [31]-[33]:

$$|Q_{gr}| = \frac{k}{2}(V_c^2 + \alpha/k) \quad (A22)$$

where $\alpha=2.\rho_0 e$ is a residual charge ($\rho_0$) related term whereas chemical potential at source and drain is calculated as [31]:

$$V_{cs,d} = \frac{C - \sqrt{C^2 \pm 2k[C_{top}(V_G - V_{GS0} - V_{S,DINT}) + C_{back}(V_B - V_{BS0} - V_{S,DINT})]}}{\pm k} \quad (A23)$$

In the denominator of (A20), $L_{eff}$ is defined which represents an effective length to take into account Velocity Saturation (VS) effect. The thorough procedure of its extraction will be presented below. To proceed with the calculation, two cases should be distinguished regarding $u_{sat}$ value as described below; one for $V_c < V_{ccrit}$ where $u_{sat}$ is constant and the other for the opposite conditions where $u_{sat}$ is inversely proportional to sqrt($V_c^2+a/k$) [33].

$$u_{sat} = \begin{cases} \frac{2v_f}{\pi} = S = 6.62 \cdot 10^5 \text{m/s} \rightarrow |V_c| < V_{ccrit} \\ \frac{\Omega}{\sqrt{\frac{\pi |Q_{gr}|}{e}}} = \frac{\Omega}{\sqrt{\frac{\pi k(V_c^2 + \frac{\alpha}{k})}{2e}}} \frac{\Omega h u_f}{e\sqrt{V_c^2 + \frac{\alpha}{k}}} = \frac{N}{\sqrt{V_c^2 + \frac{\alpha}{k}}} \quad , N = \frac{h\Omega u_f}{e} \rightarrow |V_c| > V_{ccrit} \end{cases} \quad (A24)$$

For the first case where $u_{sat}$ is constant we take:

$$L_{eff} = L + \mu \int_{V_{cs}}^{V_{cd}} \frac{k|V_c|}{SC} |dV_c| = L + \frac{\mu}{SC} \left|\left[\pm \frac{1}{2} k V_c^2\right]_{V_{cd}}^{V_{cs}}\right| \rightarrow |V_c| < V_{ccrit} \quad (A25a)$$

whereas for the second case where $u_{sat}$ is inversely proportional to sqrt($V_c^2+a/k$) we have:

$$L_{eff} = L + \mu \int_{V_{cs}}^{V_{cd}} \frac{\sqrt{V_c^2 + \frac{\alpha}{k}} k|V_c|}{NC} |dV_c| = L + \frac{\mu k}{NC} \left|\left[\pm \frac{1}{3}\left(V_c^2 + \frac{\alpha}{k}\right)^{3/2}\right]_{V_{cd}}^{V_{cs}}\right| \rightarrow |V_c| > V_{ccrit} \quad (A25b)$$

The absolute value in the analytical solution of (A25) comes from $|dV_c|$ in order to distinguish two cases for $L_{eff}$ depending on the sign of $dV_c$. Thus, in the case of $dV_c<0 \rightarrow V_{cs}>V_{cd}$ ($V_{DS}>0$) $\rightarrow |dV_c|=-dV_c$, the integral is solved from $V_{cd}$ to $V_{cs}$ while when $dV_c>0 \rightarrow V_{cs}<V_{cd}$ ($V_{DS}<0$) $\rightarrow |dV_c|=dV_c$, the integral is solved from $V_{cs}$ to $V_{cd}$.

* $\pm, \mp$ : Top sign refers to $V_c>0$ and bottom sign to $V_c<0$.

**C. Supplementary Information: Thorough procedure for channel drain current noise derivations.**

**C1. General methodology**

As described in [19 §6.1.1], the methodology for noise derivations applied here, considers a noiseless channel apart from an elementary slice between x and x+Δx as shown in Fig. S1b. This local noise contribution can be represented by a local current noise source with a Power Spectral Density (PSD) $S_{\delta I^2_n}$ which is connected in parallel with the resistance ΔR of the slice. The transistor then can be split into two noiseless transistors M1 and M2 on each side of the local current noise source, at the source and drain side ends with channel lengths equal to x and L-x respectively. Since the voltage fluctuations on parallel resistance ΔR are small enough compared to thermal voltage $U_T$, small signal analysis can be used in order to extract a noise model according to which, M1 and M2 can be replaced by two simple conductances $G_S$ on the source and $G_D$ on the drain side, respectively. The PSD of the drain current fluctuations $S_{\delta I^2_{nD}}$ due to a single local noise source is given by [19, (6.3)]:

$$S_{\delta I^2_{nD}}(\omega, x) = G^2_{CH} \Delta R^2 S_{\delta I^2_{\tilde{n}}}(\omega, x) \tag{A26}$$

where $\omega$ is the angular frequency and $G_{ch}$ is the channel conductance at x where [19, (6.2)]:

$$\frac{1}{G_{CH}} = \frac{1}{G_S} + \frac{1}{G_D} \tag{A27}$$

Total drain current noise PSD along the channel is obtained by summing the elementary contributions $S_{\delta I^2_{nD}}$ in (A26) assuming that the contribution of each slice at different positions along the channel remains uncorrelated [19, (6.4)]:

$$S_{ID} = \int_0^L \frac{S_{\delta I^2_{nD}}(\omega, x)}{\Delta x} dx = \int_0^L G^2_{CH} \Delta R^2 \frac{S_{\delta I^2_{\tilde{n}}}(\omega, x)}{\Delta x} dx \tag{A28}$$

which is also (3) of the main manuscript.

## C2. Useful relations

In this subsection, we provide the complete step-by-step derivations of (7), (10)-(12) and (14) of the main manuscript. Moreover, some useful relations between effective mobility $\mu_{eff}$, its derivative w. r. t. longitudinal electric field $\mu'_{eff}$ and differential mobility $\mu_{diff}$ are calculated which will be very helpful for the derivation of the final noise compact model. In more detail, (7) of the main manuscript is solved due to (2):

$$\frac{\partial I_D}{\partial E} = \frac{-\partial W|Q_{gr}|\mu_{eff}E}{\partial E} = -W|Q_{gr}|\mu_{eff} - W|Q_{gr}|\frac{\partial \mu_{eff}}{\partial E_x}\frac{\partial E_x}{\partial E}E = -W|Q_{gr}|(\mu_{eff} + \mu'_{eff}E_X) = -W|Q_{gr}|\mu_{diff}$$
$$\tag{A29}$$

Equation (10) of the main manuscript is solved due to (8), (9) and (A27):

$$\frac{1}{G_{CH}} = \frac{1}{G_S} + \frac{1}{G_D} = \frac{x + \int_{V_S}^V \frac{\mu'_{eff}\frac{\partial E_x}{\partial E}}{\mu_{diff}}dV}{W|Q_{gr}|\mu_{eff}} + \frac{L - x + \int_V^{V_D}\frac{\mu'_{eff}\frac{\partial E_x}{\partial E}}{\mu_{diff}}dV}{W|Q_{gr}|\mu_{eff}} = \frac{L + \int_{V_S}^{V_D}\frac{\mu'_{eff}\frac{\partial E_x}{\partial E}}{\mu_{diff}}dV}{W|Q_{gr}|\mu_{eff}} \rightarrow G_{CH} = \frac{W|Q_{gr}|\mu_{eff}}{L + \int_{V_S}^{V_D}\frac{\mu'_{eff}\frac{\partial E_x}{\partial E}}{\mu_{diff}}dV} \tag{A30}$$

Equation (11) of the main manuscript is solved due to (9), (A17) and $\mu_{diff}$ definition:

$$\Delta R = \frac{1}{\Delta G} = \frac{\Delta x + \int_x^{x+\Delta x}\frac{\mu'_{eff}\frac{\partial E_x}{\partial E}dV}{\mu_{diff}}dx}{W|Q_{gr}|\mu_{eff}} = \frac{\Delta x + \int_x^{x+\Delta x} -\frac{\mu'_{eff}\frac{\partial E_x}{\partial E}}{\mu_{diff}}Edx}{W|Q_{gr}|\mu_{eff}} = \frac{\Delta x\left(1 - E_x\frac{\mu'_{eff}}{\mu_{diff}}\right)}{W|Q_{gr}|\mu_{eff}} = \frac{\Delta x(\mu_{eff} + \mu'_{eff}E_x - \mu'_{eff}E_x)}{W|Q_{gr}|\mu_{eff}\mu_{diff}} = \frac{\Delta x}{W|Q_{gr}|\mu_{diff}}$$
$$\tag{A31}$$

Equation (12) of the main manuscript is solved due to (11) and $\Delta \tilde{N}^2/\tilde{N} = (K_B T_L/n_{gr}) \cdot (\partial n_{gr}/\partial E_F)$ where $\Delta \tilde{N}^2$ is the variance and $\tilde{N}$ the average number of carriers [29, (3)], [30]:

$$S_{\delta I_n^2}(\omega, x) = \frac{4K_B T_n}{\Delta R} = \frac{4K_B T_c}{\Delta R} \frac{\Delta \bar{N}^2}{\bar{N}} = \frac{4K_B T_c}{\Delta R} \frac{KT_L}{n_{gr}} \frac{\partial n_{gr}}{\partial EF} = \frac{4K_B T_c}{\Delta R} \frac{KT_L}{|Q_{gr}|} \frac{\frac{\partial |Q_{gr}|}{\partial V_C}}{\frac{\partial EF}{\partial V_C}} = \frac{4K_B T_c}{\Delta R} \frac{K_B T_L k}{e|Q_{gr}|} |V_c| =$$

$$4K_B T_C \frac{W \mu_{diff} U_T}{\Delta x} k |V_c| \tag{A32}$$

Regarding mobility relations, we have:

$$\mu_{eff} = \frac{\mu}{1 + \frac{|E_x|}{E_C}} = \frac{\mu E_C}{|E_x| + E_C} \leftrightarrow \mu_{eff}^2 = \frac{(\mu E_C)^2}{(|E_x| + E_C)^2} \tag{A33}$$

and

$$\mu'_{eff} = \frac{\partial \mu_{eff}}{\partial E_x} = \frac{\partial \left(\frac{\mu E_C}{|E_x| + E_C}\right)}{\partial E_x} = \frac{-\mu E_C E_x}{|E_x|(|E_x| + E_C)^2} \tag{A34}$$

Due to (A33), (A34), differential mobility is given:

$$\mu_{diff} = \mu_{eff} + \mu'_{eff} E_x = \frac{\mu E_C}{|E_x| + E_C} + \frac{-\mu E_C E_x^2}{|E_x|(|E_x| + E_C)^2} = \frac{\mu E_C |E_x|(|E_x| + E_C) - \mu E_C E_x^2}{|E_x|(|E_x| + E_C)^2} = \frac{\mu E_C^2}{(|E_x| + E_C)^2} \tag{A35}$$

and from (A33), (A35):

$$\frac{\mu_{eff}^2}{\mu_{diff}} = \frac{\frac{(\mu E_C)^2}{(|E_x| + E_C)^2}}{\frac{\mu E_C^2}{(|E_x| + E_C)^2}} = \mu \tag{A36}$$

whereas from (A34), (A35):

$$\frac{\mu'_{eff}}{\mu_{diff}} = \frac{\frac{-\mu E_C E_x}{|E_x|(|E_x| + E_C)^2}}{\frac{\mu E_C^2}{(|E_x| + E_C)^2}} = \frac{-E_x}{E_C |E_x|} = \frac{\frac{d\psi}{dx}}{E_C \left|-\frac{d\psi}{dx}\right|} = \frac{\frac{d\psi}{dV_C} dV_C}{E_C \left|-\frac{d\psi}{dV_C} dV_C\right|} = \frac{-\frac{C_q}{C} dV_C}{E_C \left|\frac{C_q}{C} dV_C\right|} = \frac{-dV_C}{E_C |dV_C|} \tag{A37}$$

Equation (14) of the main manuscript is solved due to (A17), (A20) and (A37):

$$M = \frac{1}{\left(1 + \frac{1}{L} \int_{V_S}^{V_D} \frac{\mu'_{eff}}{\mu_{diff}} \frac{\partial E_x}{\partial E} dV\right)^2} = \frac{1}{\left(1 + \frac{\mu}{L} \int_{V_{CS}}^{V_{cd}} \frac{-dV_C}{u_{sat}|dV_C|} \frac{C_q}{C_q + C} \left(-\frac{C_q + C}{C}\right) dV_C\right)^2} = \frac{1}{\left(1 + \frac{\mu}{CL} \int_{V_{CS}}^{V_{cd}} \frac{C_q dV_C}{u_{sat}|dV_C|} dV_C\right)^2} =$$

$$\frac{1}{\left(1 + \frac{\mu}{CL} \int_{V_{CS}}^{V_{cd}} \frac{C_q}{u_{sat}} |dV_C|\right)^2} = \left(\frac{L}{L_{eff}}\right)^2 \tag{A38}$$

### C3. $S_{IDB}$, $S_{IDC}$ thermal noise integrals – degenerate nature of graphene

$S_{IDB}$ after the consideration of (A17), (A19) becomes:

$$S_{IDB} = 4K_B T_L U_T k \mu \frac{W}{L_{eff}^2} \int_0^L 2 \frac{|E_x|}{E_C} |V_c| dx =$$

$$4K_B T_L U_T k \mu \frac{W}{L_{eff}^2} \int_0^L \frac{2\mu}{u_{sat}} |E||V_c| \frac{C_q}{C_q + C} dx = 4K_B T_L U_T k \mu \frac{W}{L_{eff}^2} \int_{V_S}^{V_D} \frac{2\mu}{u_{sat}} \frac{C_q}{C_q + C} |V_c||-dV| =$$

$$4K_B T_L U_T k \mu \frac{W}{L_{eff}^2} \int_{V_{CS}}^{V_{cd}} \frac{2\mu}{u_{sat}} \frac{C_q}{C_q + C} |V_c| \left|-\frac{dV}{dV_c} dV_c\right| = 4K_B T_L U_T k \mu \frac{W}{L_{eff}^2} \int_{V_{CS}}^{V_{cd}} \frac{2\mu}{u_{sat}} \frac{C_q}{C} |V_c||dV_c| \tag{A39}$$

It is apparent from (19) in the main manuscript and (A39) that $S_{IDB}$ equals to the double of $S_{IDA2}$.

$S_{IDC}$ is given by the following equation if (A17), (A19) are considered. Electric field is written as $E^2=(-dV/dx)(-dV/dx)$ and then both sides are integrated after being multiplied with dx.

$$S_{IDC} = 4K_BT_LU_Tk\mu \frac{W}{L_{eff}^2} \int_0^L \left(\frac{E_x}{E_C}\right)^2 |V_c|dx = 4K_BT_LU_Tk\mu \frac{W}{L_{eff}^2} \int_0^L \frac{\mu^2}{u_{sat}^2}\left(\frac{C_q}{C_q+C}\right)^2 |V_c|\left|-\frac{dV}{dx}\right|\left|-\frac{dV}{dx}\right| dx \Leftrightarrow$$

$$dxS_{IDC} = 4K_BT_LU_Tk\mu \frac{W}{L_{eff}^2} \int_{V_{cs}}^{V_{cd}} \frac{\mu^2}{u_{sat}^2}\left(\frac{C_q}{C_q+C}\right)^2 |V_c|\left|-\frac{dV}{dV_c}dV_c\right|\left|-\frac{dV}{dV_c}dV_c\right| \Leftrightarrow \int_0^L S_{IDC}dx =$$

$$4K_BT_LU_Tk\mu \frac{W}{L_{eff}^2} \int_{V_{cs}}^{V_{cd}} \int_{V_{cs}}^{V_{cd}} \frac{\mu^2}{u_{sat}^2}\left(\frac{C_q}{C_q+C}\right)^2 \left(\frac{C_q+C}{C}\right)^2 |V_c||dV_c||dV_c| \Leftrightarrow S_{IDC} =$$

$$4K_BT_LU_Tk\mu^3 \frac{W}{LC^2L_{eff}^2} \int_{V_{cs}}^{V_{cd}} \int_{V_{cs}}^{V_{cd}} \frac{(k|V_c|)^3}{u_{sat}^2} |dV_c||dV_c| \qquad (A40)$$

**C4. Thermal noise integrals – non-degenerate approximation**

The thermal noise integrals in (18)-(20) of the main manuscript are derived and solved after the degenerate nature of graphene is considered. Since all of the analytical models available in literature consider a non-degenerate approximation [19], thermal noise for non-degenerate case should also be calculated for graphene for comparison reasons. Thus, the local noise source PSD is also calculated for the non-degenerate case where it is given:

$$S_{\delta I_n^2}(\omega, x) = \frac{4K_BT_n}{\Delta R} = 4K_BT_C \frac{W\mu_{diff}|Q_{gr}|}{\Delta x} \qquad (A41)$$

If (A41) is considered for the local noise source, then total drain current noise in (16) of the main manuscript is calculated as:

$$S_{ID-ND} = 4K_BT_L\mu \frac{W}{L_{eff}^2}\left[\int_0^L |Q_{gr}|dx + \int_0^L 2\frac{|E_x|}{E_C}|Q_{gr}|dx + \int_0^L \left(\frac{E_x}{E_C}\right)^2 |Q_{gr}|dx\right] \qquad (A42)$$

As in degenerate case, (A42) can be split into 3 integrals named $S_{IDA-ND}$, $S_{IDB-ND}$, $S_{IDC-ND}$, respectively. In order to solve each one of them, the integral variable change from x to $V_c$ described in (A19) shall be applied. More specifically for $S_{IDA-ND}$:

$$S_{IDA-ND} = 4K_BT_Lk\mu \frac{W}{L_{eff}^2} \int_0^L |Q_{gr}|dx = 4K_BT_L\mu \frac{W}{L_{eff}^2}\left[\begin{array}{c}\int_{V_{cs}}^{V_{cd}}\left(-\frac{|Q_{gr}|^2 2L_{eff}}{kg_{vc}}\left(\frac{C_q+C}{C}\right)\right)dV_C - \\ \int_{V_{cs}}^{V_{cd}}\left(\frac{\mu|Q_{gr}|}{v_{sat}}\left(\frac{C_q}{C}\right)\right)|dV_c|\end{array}\right] \qquad (A43)$$

which again is split into two integrals, namely $S_{IDA1-ND}$ (1st in the brackets) and $S_{IDA2-ND}$ (2nd in the brackets) as $S_{IDA-ND} = S_{IDA1-ND} - S_{IDA2-ND}$ where:

$$S_{IDA1-ND} = 2K_BT_Lk\mu \frac{W}{C_{gvc}L_{eff}} \int_{V_{cd}}^{V_{cs}}\left(\left(V_c^2 + \frac{\alpha}{k}\right)^2 (k|V_c| + C)\right)dV_C = 2K_BT_Lk\mu \frac{W}{C_{gvc}L_{eff}}\left[\frac{\alpha^2 CV_c}{k^2} \pm \frac{\alpha^2 V_c^2}{2k} + \right.$$

$$\left. \frac{2\alpha CV_c^3}{3k} \pm \frac{\alpha V_c^4}{2} + \frac{CV_c^5}{5} \pm \frac{kV_c^6}{6}\right]_{V_{cd}}^{V_{cs}} \qquad (A44)$$

while $S_{ID2\text{-}ND}$ depends on $u_{sat}$ model described in (A24), thus two different cases shall be considered: Near CNP:

$$S_{IDA2-ND} = 2K_BT_Lk\mu^2 \frac{W}{CL_{eff}^2} \int_{V_{cs}}^{V_{cd}} \left(\frac{\left(V_c^2+\frac{\alpha}{k}\right)^2 k|V_c|}{v_{sat}}\right) |dV_c| = 2K_BT_Lk\mu^2 \frac{W}{CSL_{eff}^2} \left|\left[\pm\frac{\alpha V_c^2}{2} \pm \frac{kV_c^4}{4}\right]_{V_{cd}}^{V_{cs}}\right| \rightarrow |V_c| <$$

$V_{ccrit}$ (A45a)

and away CNP:

$$S_{IDA2-ND} = 2K_BT_Lk\mu^2 \frac{W}{CL_{eff}^2} \int_{V_{cs}}^{V_{cd}} \left(\frac{\left(V_c^2+\frac{\alpha}{k}\right)^{3/2} k|V_c|}{N}\right) |dV_c| = 2K_BT_Lk\mu^2 \frac{W}{CNL_{eff}^2} \left|\left[\frac{\pm k}{5}(V_c^2 + \alpha/k)^{5/2}\right]_{V_{cd}}^{V_{cs}}\right| \rightarrow$$

$|V_c| > V_{ccrit}$ (A45b)

The absolute value in the analytical solution of (A45) comes from $|dV_c|$ in order to distinguish two cases for $S_{IDA2\text{-}ND}$ depending on the sign of $dV_c$. Thus, in the case of $dV_c<0 \rightarrow V_{cs}>V_{cd}$ ($V_{DS}>0$) $\rightarrow |dV_c|=-dV_c$, the integral is solved from $V_{cd}$ to $V_{cs}$ while when $dV_c>0 \rightarrow V_{cs}<V_{cd}$ ($V_{DS}<0$) $\rightarrow |dV_c|=dV_c$, the integral is solved from $V_{cs}$ to $V_{cd}$. $S_{IDB\text{-}ND}$ after the consideration of (A17), (A19) becomes:

$$S_{IDB-ND} = 4K_BT_L\mu \frac{W}{L_{eff}^2} \int_0^L 2\frac{|E_x|}{E_C}|Q_{gr}|dx = 2K_BT_Lk\mu \frac{W}{L_{eff}^2} \int_0^L \frac{2\mu}{u_{sat}}|E|(V_c^2 + \alpha/$$

$$k)\frac{C_q}{C_q+C}dx = 2K_BT_Lk\mu \frac{W}{L_{eff}^2} \int_{V_S}^{V_D} \frac{2\mu}{u_{sat}} \frac{C_q}{C_q+C}(V_c^2 + \alpha/k)|-dV| = 2K_BT_Lk\mu \frac{W}{L_{eff}^2} \int_{V_{cs}}^{V_{cd}} \frac{2\mu}{u_{sat}} \frac{C_q}{C_q+C}(V_c^2 + \alpha/$$

$$k)\left|-\frac{dV}{dV_c}dV_c\right| = 2K_BT_Lk\mu \frac{W}{L_{eff}^2} \int_{V_{cs}}^{V_{cd}} \frac{2\mu}{u_{sat}} \frac{C_q}{C}(V_c^2 + \alpha/k)|dV_c|$$ (A46)

It is apparent from (A43), (A46) that $S_{IDB\text{-}ND}$ equals to the double of $S_{IDA2\text{-}ND}$. $S_{IDC\text{-}ND}$ is given by the following equation if (A17), (A19) are considered. Electric field is written as $E^2=(-dV/dx)(-dV/dx)$ and then both sides are integrated after being multiplied with $dx$.

$$S_{IDC-ND} = 4K_BT_L\mu \frac{W}{L_{eff}^2} \int_0^L \left(\frac{E_x}{E_C}\right)^2 |Q_{gr}|dx = 2K_BT_Lk\mu \frac{W}{L_{eff}^2} \int_0^L \frac{\mu^2}{u_{sat}^2}\left(\frac{C_q}{C_q+C}\right)^2 (V_c^2 + \alpha/k)\left|-\frac{dV}{dx}\right|\left|-\frac{dV}{dx}\right|dx \Leftrightarrow$$

$$dxS_{IDC-ND} = 2K_BT_Lk\mu \frac{W}{L_{eff}^2} \int_{V_{cs}}^{V_{cd}} \frac{\mu^2}{u_{sat}^2}\left(\frac{C_q}{C_q+C}\right)^2 (V_c^2 + \alpha/k)\left|-\frac{dV}{dV_c}dV_c\right|\left|-\frac{dV}{dV_c}dV_c\right| \Leftrightarrow \int_0^L S_{IDC-ND}dx =$$

$$2K_BT_Lk\mu \frac{W}{L_{eff}^2} \int_{V_{cs}}^{V_{cd}}\int_{V_{cs}}^{V_{cd}} \frac{\mu^2}{u_{sat}^2}\left(\frac{C_q}{C_q+C}\right)^2 \left(\frac{C_q+C}{C}\right)^2 (V_c^2 + \alpha/k)|dV_c||dV_c| \Leftrightarrow S_{IDC-ND} =$$

$$2K_BT_Lk\mu^3 \frac{W}{LC^2L_{eff}^2} \int_{V_{cs}}^{V_{cd}}\int_{V_{cs}}^{V_{cd}} \frac{(kV_c)^2}{u_{sat}^2}(V_c^2 + \alpha/k)|dV_c||dV_c|$$ (A47)

Again, $S_{IDC\text{-}ND}$ depends on $u_{sat}$ model described in (A24), thus two different cases shall be considered: Near CNP:

$$S_{IDC-ND} = 2K_B T_L k\mu^3 \frac{W}{LC^2 L_{eff}^2} \int_{V_{cs}}^{V_{cd}} \int_{V_{cd}}^{V_{cs}} \frac{(kV_c)^2}{S^2}\left(V_c^2 + \frac{\alpha}{k}\right) |dV_c||dV_c| = 2K_B T_L k\mu^3 \frac{W}{LS^2 C^2 L_{eff}^2}(V_{cs} - V_{cd}) \left[\frac{k\alpha}{3}V_c^3 + \frac{k^2 V_c^5}{5}\right]_{V_{cd}}^{V_{cs}} \to |V_c| < V_{ccrit} \qquad (A48a)$$

and away CNP:

$$S_{IDC-ND} = 2K_B T_L k\mu^3 \frac{W}{LC^2 L_{eff}^2} \int_{V_{cs}}^{V_{cd}} \int_{V_{cd}}^{V_{cs}} \frac{(kV_c)^2}{N^2}\left(V_c^2 + \frac{\alpha}{k}\right)^2 |dV_c||dV_c| = 2K_B T_L k\mu^3 \frac{W}{LN^2 C^2 L_{eff}^2}(V_{cs} - V_{cd}) \left[\frac{\alpha^2}{3}V_c^3 + \frac{2\alpha k}{5}V_c^5 + \frac{k^2 V_c^7}{7}\right]_{V_{cd}}^{V_{cs}} \to |V_c| > V_{ccrit} \qquad (A48b)$$

It is clear from (A48) that $S_{IDC-ND}$ has always the same solution since the sign of the product $|dV_c||dV_c|=dV_c.dV_c$ for $dV_c>0$ ($V_{DS}<0$) or $|dV_c||dV_c|=(-dV_c)(-dV_c)$ for $dV_c<0$ ($V_{DS}>0$) is always positive..
As mentioned before $S_{ID-ND}= S_{IDA-ND} + S_{IDB-ND} + S_{IDC-ND} = S_{IDA1-ND} + S_{IDA2-ND} + S_{IDC-ND}$.

* $\pm, \mp$ : Top sign refers to $V_c>0$ and bottom sign to $V_c<0$.